\documentclass[a4paper,12pt]{article}

\usepackage[margin=2.5cm]{geometry}
\usepackage{authblk}
\usepackage{graphicx}
\usepackage{caption}
\usepackage{mathptmx}
\usepackage{amsmath}
\usepackage{amsfonts}
\usepackage[
    colorlinks=true,
    linkcolor=red,
    citecolor=blue,
    breaklinks=true
]{hyperref}

\title{Resolving the Topological Classification of Bismuth with Topological Defects}

\author[1]{Abhay Kumar Nayak$^\dagger$}
\author[1]{Jonathan Reiner$^\dagger$}
\author[1]{Raquel Queiroz}
\author[1]{Huixia Fu}
\author[2]{Chandra Shekhar}
\author[1]{Binghai Yan}
\author[2]{Claudia Felser}
\author[1]{Nurit Avraham}
\author[1]{\authorcr Haim Beidenkopf\thanks{email: haim.beidenkopf@weizmann.ac.il, $^\dagger$These authors contributed equally to this work.}}
\affil[1]{Department of Condensed Matter Physics, Weizmann Institute of Science, Rehovot 7610001, Israel}
\affil[2]{Max Planck Institute for Chemical Physics of Solids, D-01187 Dresden, Germany}
\date{}

\begin{document}

\maketitle
\captionsetup[figure]{labelfont={bf},labelformat={default},labelsep=period}

\begin{abstract}

Bulk boundary correspondence in topological materials allows to study their bulk topology through the investigation of their topological boundary modes \cite{Kane2005, Bernevig2006, Fu2007, Hasan2010, Qi2011, BERNEVIG2013, Bansil2016, Wen2017, Armitage2018}. However, for classes that share similar boundary phenomenology, the growing diversity of topological phases may lead to ambiguity in the topological classification of materials. Such is the current status of bulk bismuth. While some theoretical models indicate that bismuth possesses a trivial topological nature, other theoretical and experimental studies suggest non-trivial topological classifications such as a strong or a higher order topological insulator, both of which hosts helical modes on their boundaries \cite{Liu1995, Fu2007, Teo2008, Ohtsubo2016, Schindler2018a}. Here we use a novel approach to resolve the topological classification of bismuth by spectroscopically mapping the response of its boundary modes to a topological defect in the form of a screw dislocation (SD). We find that the edge mode extends over a wide energy range, and withstands crystallographic irregularities, without showing any signs of backscattering. It seems to bind to the bulk SD, as expected for a topological insulator (TI) with non-vanishing weak indices \cite{Ran2009,Teo2010a,Teo2017}. We argue that the small scale of the bulk energy gap, at the time reversal symmetric momentum $L$, positions bismuth within the critical region of a topological phase transition to a strong TI with non-vanishing weak indices. We show that the observed boundary modes are approximately helical already on the $\mathbb{Z}_2$ trivial side of the topological phase transition. This work opens the door for further possibilities to examine the response of topological phases to crystallographic topological defects, and to uniquely explore their associated bulk boundary phenomena. 
\end{abstract}

\subsection*{Introduction}

Topological materials host unique boundary states that cannot be realized as standalone electronic systems but only on the boundaries of a topologically classified bulk. These include the chiral edge modes of the quantum Hall state, the helical modes on the surface of topological insulators (TIs), Fermi-arc bands on the surface of a Weyl semimetal and Majorana modes in topological superconductors \cite{Kane2005, Bernevig2006, Fu2007, Hasan2010, Qi2011, BERNEVIG2013, Bansil2016, Wen2017, Armitage2018}. Crystalline topological defects, as dislocations and disclinations, may also serve as effective boundaries and thus bind such topologically protected boundary modes \cite{Ran2009,Teo2010,Barkeshli2012,Hughes2013,Teo2017}. Soon after the discovery of the symmetry protected topological phases a general theory of topological defects and their associated gapless modes has been developed \cite{Teo2010a}. Examples of such bound states are the Majorana modes at vortex ends in chiral superconductors \cite{Read2000, Ivanov2001} and one-dimensional (1D) helical modes at screw dislocations (SD) in topological insulators (TIs) \cite{Ran2009}. Interestingly, bulk topological defects, when terminated by a surface, give rise to unique crystallographic structures on that surface. An example of such 'structural' surface-bulk correspondence is the surface step edge that emanates from the termination point of a bulk SD on the material surface. The only way to realize a step edge that terminates at a point on the surface, away from the surface boundaries, is by its continuation into the bulk in the form of a SD (Fig.\ref{fig1}b). The height of the SD winding then dictates the height of its corresponding step-edge \cite{Kelly2012}. Here we use these crystallographic step-edges present on the surface of Bi(111) by topological SDs, to investigate the topological nature of the electronic boundary modes. The correspondence between the structure of the topological defect and the topology class of its associated protected electronic mode provides insight on its topological origin \cite{Teo2010a,Teo2017}. Using scanning tunneling microscopy (STM) and spectroscopic mappings we investigate the local response of the various boundary electronic states on the (111) surface of  bismuth (Bi) to the surface perturbations imposed by topological SDs.    

Bi, an extensively studied low-density semi-metal, has a rich history of intriguing quantum phenomena like electron fractionalization \cite{Behnia2007}, nematic quantum Hall phases \cite{Feldman2016,Randeria2018} and unconventional superconductivity \cite{Prakash2017}. Due to its large spin-orbit coupling it plays a fundamental role in several topological materials, such as the first realized three dimensional (3D) strong TI, Bi$_{1-x}$Sb$_x$ \cite{Fu2007, Murakami2007, Hsieh2008, Roushan2009}, that shows almost identical 3D band structure to that of pure Bi \cite{Benia2015}, as well as the canonical TIs Bi$_2$Se$_3$ and Bi$_2$Te$_3$ that host a single Dirac surface band \cite{Zhang2009a, Chen2009, Hsieh2009a, Xia2009}. Nevertheless, the topological classification of pure Bi has remained, thus far, rather ambiguous. Many theoretical models \cite{Fu2007,Teo2008,Zhang2009,Liu1995} indicate that Bi possesses a trivial $\mathbb{Z}_2$ topological classification. Since bulk Bi is inversion symmetric, its strong topological index, $\nu_0$, and the weak indices $\mathbf{\nu}=(\nu_1, \nu_2, \nu_3)$ can be calculated from the parity eigenvalues of the occupied bands at time-reversal invariant momentum (TRIM) points ($\Gamma$, $T$, and three inequivalent $L$ and $X$ points, shown in Fig.\ref{fig1}a) \cite{Fu2007, Teo2008}. According to this, both the strong and weak indices of Bi are zero. 

Nevertheless, several studies demonstrate phenomenology which implies non-trivial topology in Bi, both in bulk crystals, thin films as well as in its two dimensional (2D) limit of a bilayer \cite{Yang2012, Ohtsubo2013, Drozdov2014, Peng2018a, Schindler2018a}. Direct visualization of the electronic structure of Bi around the $\bar{M}$ surface energy gap \cite{Ito2016}, suggests the 2D surface bands may have topological origin. A bilayer of Bi was among the first materials to be classified as a 2D TI \cite{Murakami2006}. Later, 1D edge modes were observed in thin films of Bi \cite{Kawakami2015, Takayama2015, Liu2017a, Peng2018a}, as well as on step edges of cleaved (111) surfaces of bulk Bi crystals \cite{Drozdov2014}. Various interpretations have been ascribed to the step-edge 1D edge modes. These include trivial spin-split bands of the 1D Rashba type \cite{Yeom2016}, helical edge modes of a 2D TI \cite{Drozdov2014}, and more recently, helical hinge modes of a 3D higher order TI \cite{Schindler2018a}. Helical 1D edge modes are counter propagating time-reversed partners of each other which form a Kramer's pair. As a consequence, it is impossible to introduce any backscattering between the two states, unless time-reversal symmetry is broken. This is the origin of topological protection of helical edge states. We show here, by the aid of SDs, the robust behavior of the 1D edge mode in Bi both at the SD as well as along the step that emanates from it. This alludes towards a weak TI classification of Bi rather than a higher order TI. We show that the topologically non trivial phenomenology of the 1D edge mode is inherent to the existence of an extremely small energy gap of about $5-15$ meV at the $L$ TRIM point \cite{Liu1995} (Fig.\ref{fig1}c). We argue that this energy gap classifies Bi as a $\mathbb{Z}_2$ trivial semimetal (or a higher order TI) on the verge of a topological phase transition into a strong TI with non-zero weak indices. We show that protected 1D edge modes, on the step edge and the SD, appears on both sides of the topological phase transition \cite{Xu2015d, Barriga2018}.

\subsection*{Results}

\begin{figure}[ht]
\centering
\includegraphics[width=1\textwidth]{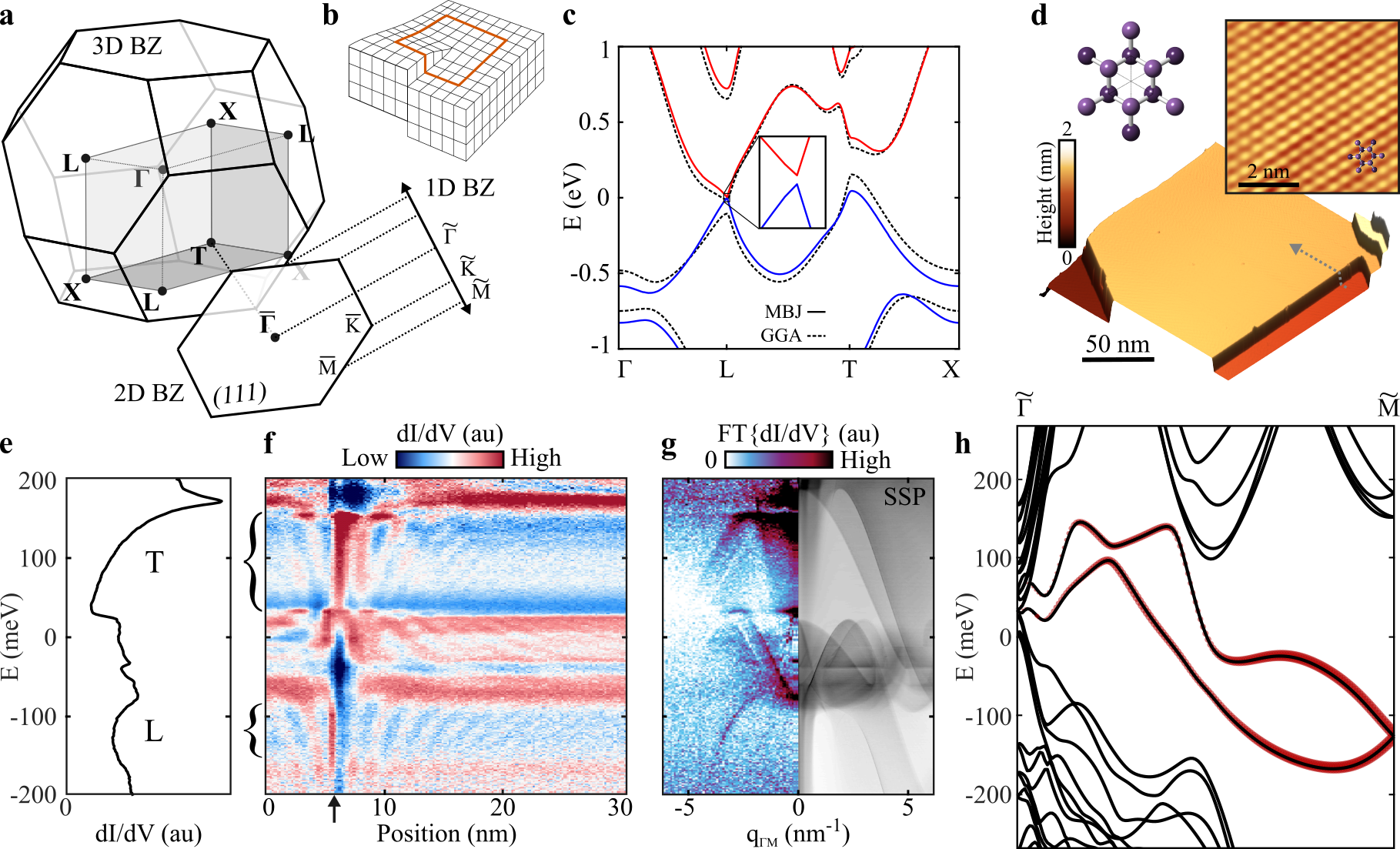}
\caption{Surface and edge spectrum of Bi (111) \textbf{a,} Bulk Brillouin Zone (BZ) of rhombohedral Bi and its (111) surface projected 2D BZ and zigzag edge projected 1D BZ, with all TRIM points indicated. \textbf{b,} Schematic illustration of a screw dislocation in a lattice. \textbf{c,} Bulk band structure (trivial $\mathbb{Z}_2$) of Bi obtained from \textit{ab initio} calculations using the modified Becke-Johnson (MBJ) exchange potential and generalized gradient approximation shown by solid and dashed lines, respectively. Inset shows the band structure around the energy gap (5 meV) at $L$, calculated with MBJ. \textbf{d,} Large scale STM topography of a Bi terrace with crystallographic step edges. Inset shows atomically resolved image, overlaid with Bi bilayer crystal structure. \textbf{e,} dI/dV spectroscopy measured at a point on the surface far away from impurities. \textbf{f,} dI/dV map measured normal to the step edge (marked with a black arrow) along the gray dotted line in \textbf{d} showing the modulated LDOS and the one-dimensional topological edge mode (marked by curly brackets).  \textbf{g,} Fourier transform of \textbf{f} showing the energy dispersion of scattering wave-vectors along $\overline{\Gamma}-\overline{M}$ and the corresponding calculated spin-selective scattering probability. \textbf{h,} Calculated zigzag edge projected LDOS showing 1D edge states dispersing along $\tilde{\Gamma}-\tilde{M}$ (see Supplementary Fig. \ref{fig:Bi_ribbon}).}
\label{fig1}
\end{figure}

We measured Bi samples derived from crystals of GdPtBi grown in Bi flux (see section \ref{sec:material} in the Supplementary Information) \cite{Shekhar2018}. We have found that cleaving such crystals under ultra high vacuum conditions, repeatedly exposes surfaces of residual crystalline Bi, due to its anisotropic rhombohedral structure. The slight lattice mismatch with GdPtBi lends the opportunity to investigate the properties of bulk Bi in the vicinity of SDs. Nevertheless, apart from the nucleation of isolated SDs, the Bi atomic surface structure and band structure we measured was found to be rather unaffected by the GdPtBi substrate (see Supplementary section \ref{sec:GdPtBi_mag}). The samples were measured at 4.2 K in a commercial Unisoku STM. A representative topographic image of a freshly cleaved surface, consisting of a flat terrace terminated by step edges along high symmetry crystallographic directions, is shown in Fig. \ref{fig1}d. The atomically resolved topographic image shows the underlying triangular lattice indicative of a pristine (111) cleave plane of Bi, thus allowing its spectroscopic studies. 

A typical differential conductance (dI/dV) spectrum measured on the terrace, away from any scatterers, is shown in Fig. \ref{fig1}e. The various peaks in the dI/dV spectrum represent the extrema of the 2D surface bands expected on Bi(111) \cite{Drozdov2014}. The intermediate dips around 100 meV and -100 meV correspond to a suppressed local density of states (LDOS) within the bulk gaps, located at the $T$ and $L$ TRIM points, respectively. To resolve some of these surface bands we have measured the dI/dV map, shown in Fig. \ref{fig1}f, taken normal to the step edge (along the gray dashed line in Fig. \ref{fig1}d). It shows strong spatial modulations that disperse in energy. These modulations are quasi-particle interference (QPI) patterns the electrons embed in the LDOS due to scattering off the step edge. Its Fourier analysis (Fig. \ref{fig1}g, left panel) reveals clear dispersing modes that agree well with the calculated spin-selective scattering probability, shown on the right panel. This detailed comparison indicates that the dispersing QPI patterns originate from the 2D surface bands that disperse between the $\overline{\Gamma}$ and $\overline{M}$ surface TRIM points in full agreement with previous studies of the (111) surface of bulk Bi crystals \cite{Ast2001,Drozdov2014}. 

We next investigate the 1D edge mode on the step edge, and its topological origin. Amidst all the energy dispersing QPI modes in Fig. \ref{fig1}f, a non-dispersing increased LDOS is bound along the step edge. This 1D bound state spans the energy interval of both the $T$ bulk-gap at 30-180 meV and the $L$ bulk-gap at lower energies around -100 meV (marked with curly brackets). This extended energy dispersion is consistent with our \textit{ab initio} calculations showing (Fig. \ref{fig1}h) that the edge mode indeed disperses from $\widetilde{\Gamma}$ to $\widetilde{M}$ (the edge projections of the bulk $T$ and $L$ TRIM points, respectively, see Fig.\ref{fig1}a), and extends throughout the -200 to 200 meV energy range. The topological nature of such 1D edge mode in Bi has been discussed in several previous reports \cite{Yang2012, Drozdov2014, Kim2014, Liu2017a, Schindler2018a}, where various distinct origins have been ascribed to its formation including 2D TI \cite{Drozdov2014}, higher order TI \cite{Schindler2018a} and non-topological Rashba metal \cite{Yeom2016}. However, all of these accounted for the appearance of the helical edge mode within the $T$ gap alone and disregarded its extended energy dispersion down to lower energies where the $L$ energy-gap resides. This is true also for the energy dispersion of the 2D surface modes in Bi(111). These energy and momentum extents suggests that the $L$ energy gap has an important role in the topological classification of Bi. 

\begin{figure}[ht]
\centering
\includegraphics{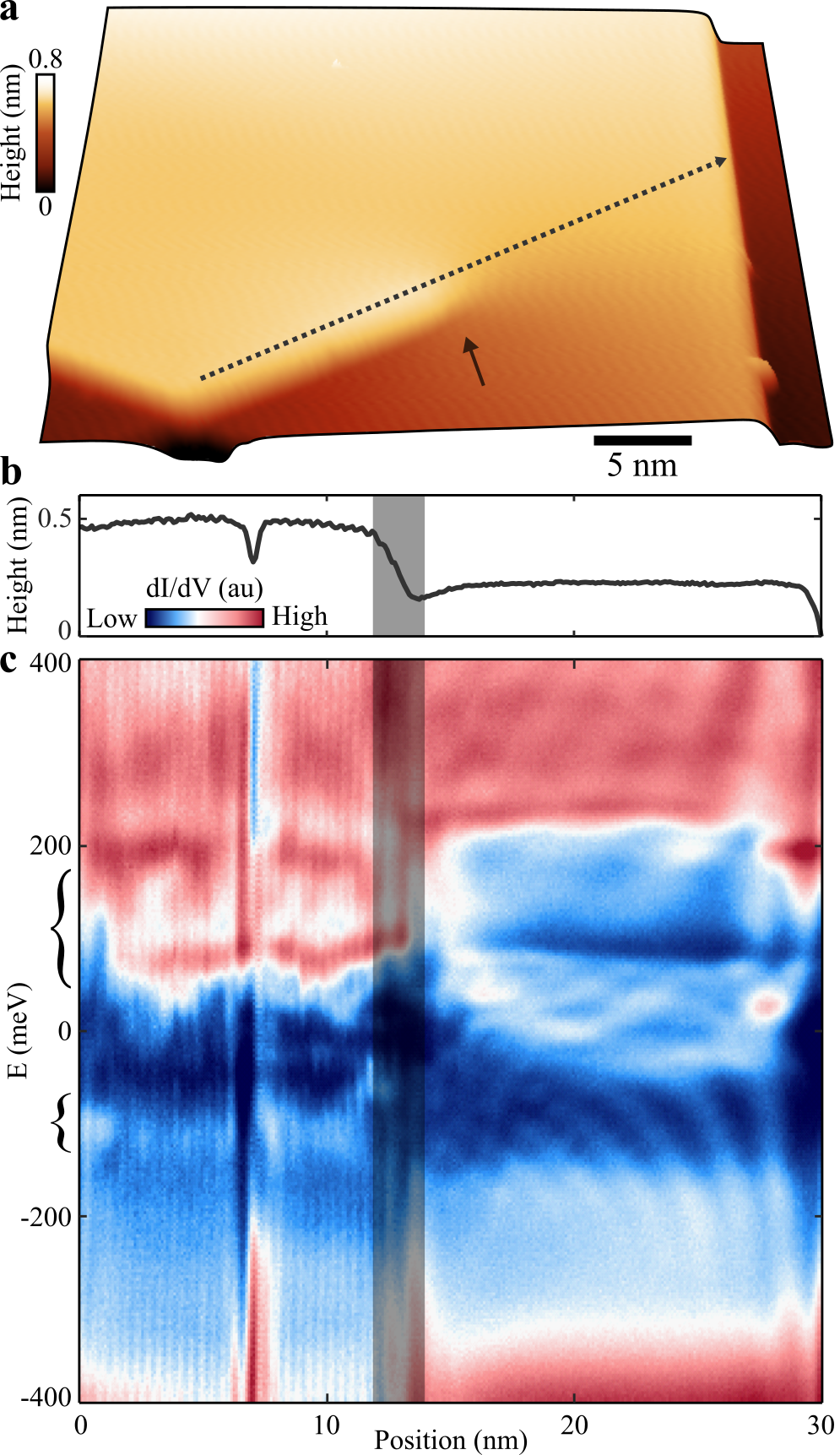}
\caption{Topographic and spectroscopic mapping of a screw dislocation. \textbf{a,} Topographic image of a clean (111) surface, consisting of a SD (solid black arrow) with a single Bi bilayer long Burger's vector. \textbf{b,} Topographic line cut along the black dotted line shown in \textbf{a}. \textbf{c,} Spectroscopic map measured along the dotted line in \textbf{a} showing a nearly non-affected edge mode on the step edge terminating at the screw dislocation. The segment where the STM tip descends from top to bottom terraces where the tunneling matrix element from tip to sample may change is shadowed.}
\label{fig2}
\end{figure}

To resolve the topological origin of the extended edge mode in Bi we examine its behavior at crystallographic defects. A topographic image of a single Bi-bilayer high step edge, terminating at a SD on the Bi(111) surface, is shown in Fig. \ref{fig2}a. The height of the step edge associated with a SD can generally be of several unit cells, $n\mathbf{u}$. This height reflects the SD Burger’s vector, $\mathbf{b}=n\mathbf{u}$, that classifies the topological invariant of the defect. We map the dI/dV spectrum along the last atomic row of the step edge, across the SD and onwards, extending into the terrace (along the dotted line in Fig. \ref{fig2}a). An atomically resolved topographic profile along this line cut, together with its corresponding dI/dV spectrum, are shown in Fig. \ref{fig2}b and c, respectively. The topographic dip signature (located about 7 nm from the left end) signifies a missing atom on the step edge. Both this vacancy, and even more so the SD, serve as scatterers to the impinging electrons. On the open terrace we find a faint quantized QPI pattern marking that both the step edge on the right and the SD indeed serve as strong scatterers for the 2D surface electrons (see also Fig. \ref{fig:LCscrew}). In sharp contrast, the 1D edge mode, which resides along the step edge (left of the SD in Fig. \ref{fig2}c), remains unperturbed all the way to the SD, without showing any signature of a gap due to backscattering off the vacancy or the SD. 

\begin{figure}[ht]
\centering
\includegraphics[scale=1]{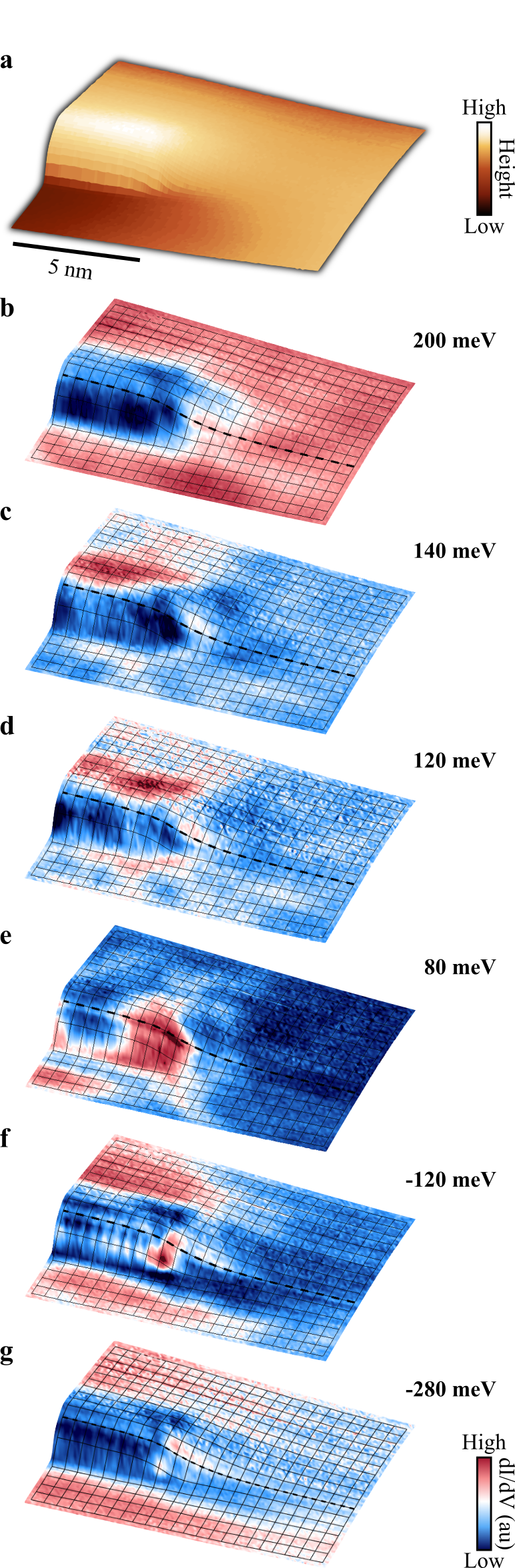}
\caption{Imaging the localization of the edge mode in the vicinity of the SD. \textbf{a,} Topographic image of the region around a SD. \textbf{b-g,} Spatially resolved differential conductance map in the vicinity of the SD. The black dashed line represents the position of the line cut shown in Fig. \textbf{\ref{fig2}c}. }
\label{fig3}
\end{figure}

To further examine the spatial structure and propagation of the edge mode near the SD as a function of energy, we overlay on a topographic height map in Fig.\ref{fig3}a 2D dI/dV mappings of the energy resolved LDOS at the SD vicinity (see also Fig. \ref{fig:SDmap}). At high energies (Fig.\ref{fig3}b) we find metallic surfaces on the SD, step edge and adjacent terraces. Below 180 meV (Fig.\ref{fig3}c-e) we detect the edge mode localized on the edge of the top terrace, all across the step edge, down to the SD. Right on the SD we observe a slight increase of the LDOS which agrees with the continuation of the mode into the SD. We observe similar behavior also at lower energies around -100 meV, shown in Fig.\ref{fig3}f, where we have identified the lower portion of the edge mode. At these energies it seems to be localized slightly lower on the step edge. Eventually, below about -200 meV the step edge becomes again less metallic than the terraces surrounding it (Fig.\ref{fig3}g). 

Absence of backscattering and hybridization of the 1D edge mode along the step edge followed by its apparent channeling into the SD seems to reflect the phenomenology of a weak TI rather than of a higher order TI. For a material characterized by non vanishing weak indices $\nu=(\nu_1,\nu_2,\nu_3)$, a bulk SD with Burger's vector \textbf{b}, is expected theoretically to bind a helical mode if $\mathbf{\nu} \cdot \mathbf{b} =\pi \pmod{2\pi}$ \cite{Ran2009, Teo2010}. The expected phenomenology on a single unit cell height step-edge, terminated by a SD, is of a 1D helical mode that runs along the step-edge and continues into the SD line, with no back scattering and gapping, as guaranteed by time reversal symmetry\cite{Ran2009, Teo2010}. This is consistent with our observations. In contrast, a higher order TI is characterized by vanishing strong index, $\nu_0=0$, and weak indices $\mathbf{\nu}=(0,0,0)$. Hence, a SD in a higher order TI is not expected to bind a helical mode, unless, as was shown recently, its Burger's vector is of a fraction of a unit cell height \cite{Queiroz2018}. Therefore, the expected phenomenology in the case of a higher order TI, is of even number of helical modes on the step-edge that would presumably gap out, or show some signatures of back scattering near the impurity and the SD. Even if the pair of hinge modes avoids hybridization \textit{along} the step edge, the SD would necessarily scatter them one to another and lead to their gradual gapping in its vicinity. The absence of such signatures, and moreover the existence of a slight increase in the LDOS on the SD, favor their identification as weak TI helical modes rather than as higher order TI hinge states. This, however, does not rule out the possibility of observing higher order TI hinge states in Bi on higher step-edges, where the upper and lower hinge modes will be further away. 

While the phenomenology in the vicinity of the SD strongly suggests the existence of non-vanishing weak indices, this is inconsistent with the majority of theoretical models of Bi \cite{Liu1995, Fu2007, Teo2008, Schindler2018a}, including our \textit{ab initio} calculations. These find the same band ordering at all TRIM points in pure Bi, leading to a trivial $\mathbb{Z}_2$ topological indices, $\nu_0=0, \mathbf{\nu}= (0,0,0)$. These conclusions, however, rely on the $L$ energy gap that in calculation is found to be as small as 5 meV (Fig. \ref{fig1}c). This small energy gap makes the determination of the band ordering at the $L$ point inconclusive. Indeed, more recent studies suggest that this energy gap is in fact inverted with respect to the energy gap at the $T$ TRIM point, which classifies Bi as a strong TI, $\nu_0$=1, with non-zero weak indices, $\mathbf{\nu}=(1,1,1)$ \cite{Ohtsubo2013, Ohtsubo2016, Aguilera2015}. In this case, the weak TI helical edge modes are expected to bind to SDs, and the 2D surface states of Bi are identified with the Dirac surface band of a strong TI, providing a compelling unified resolution to the combined boundary phenomenology observed in Bi. Such a description is potentially promoted by the strain field induced by SDs, that we address below.

\begin{figure}[ht]
\centering
\includegraphics[width=1\textwidth]{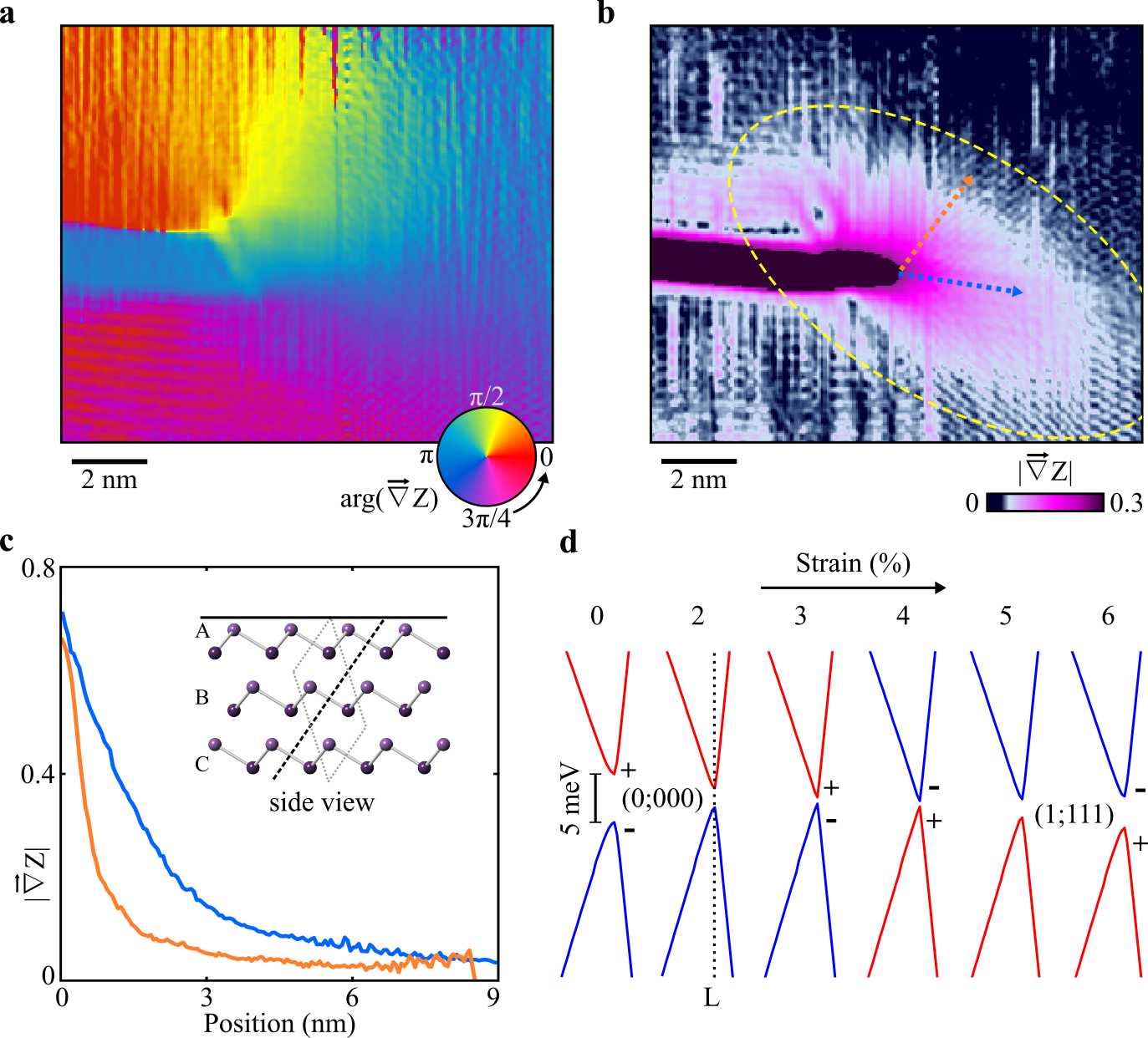}
\caption{Strain analysis in the vicinity of the SD and modeling. \textbf{a,} Phase and \textbf{b,} magnitude of the shear strain field around a SD calculated from the gradient of the topography in Fig. \textbf{\ref{fig3}a}, showing its azimuthal radially decaying profile. The amount of strain within the dashed line in \textbf{b} is sufficient to invert the band ordering around the $L$ energy gap and induce a topological phase. \textbf{c,} Radial cuts along different directions from the SD of the strain magnitude. Inset shows the (001) direction of the Burger's vector piercing the ABC stacked Bi  bilayers. \textbf{d,} \textit{ab initio} calculation of the band structure evolution around the $L$ gap under uniform strain. A topological phase transition occurs at about 3.5 \% strain.}
\label{fig4}
\end{figure}

We extract the strain field induced by the SD directly from the gradient of the measured topography, $\mathbf{\nabla} z(x,y)$. Its phase, $\arg(\mathbf{\nabla} z)$, shown in Fig.\ref{fig4}a, confirms the azimuthally oriented strain profile that winds around the SD axis. Its magnitude, $|\mathbf{\nabla} z|$, displayed in Fig.\ref{fig4}b, shows a radially decaying strain profile as the displacement of the lattice, $\mathbf{b}$, distributes over growing circumference. Cuts of the strain magnitude along the two directions marked with arrows in Fig.\ref{fig4}c, demonstrate the anisotropy of the strain distribution. 90 \% of the strain relaxes along the direction of the step-edge within 8 nm (blue arrow), while $60^\circ$ away from this direction (orange arrow), it relaxes within only  $\sim 3$ nm. This anisotropic strain pattern on the surface suggests that the SD axis is tilted away from the $<111>$ direction normal to the surface. Indeed, a single bilayer long Burger's vector along the $<111>$ direction is incompatible with the ABC stacking of the Bi bilayers (Fig. \ref{fig4}c, inset). A surface normal SD would thus result in excess strain and would be energetically unfavorable. The two observations are consistently resolved by asserting a SD oriented with the $<001>$ direction, along which the periodicity is indeed of a single bilayer. Such a SD is also expected to bind a helical mode as $\mathbf{\nu}\cdot\mathbf{b}=\pi$.

Intriguingly, our \textit{ab initio} calculations for the response of Bi to a uniform strain (Fig. \ref{fig4}d) shows that a small amount of strain is sufficient to result in band crossing at the $L$ TRIM point \cite{Hirahara2012, Ohtsubo2016} that would induce a topological phase transition into a strong TI, $\nu_0=1$, with non-zero weak indices $\mathbf{\nu}=(111)$ (see Supplementary  Fig. \ref{fig:DFTstrain} for more details). The $3-4 \%$ strain needed for topological phase transition is a small fraction of the shear strain applied at the vicinity of the SD. This suggests that the bulk topological defect would induce around it a cylindrical volume of TI nature embedded within the $\mathbb{Z}_2$ trivial bulk, as marked by the dashed line in Fig. \ref{fig4}b. We also note that recent experiments on single crystals of Bi seem to identify valley symmetry breaking due to spurious strain fields in the samples \cite{Feldman2016} which alone may be sufficient to render Bi topological throughout the bulk.

We argue that the observed phenomenology on all boundaries of Bi including surfaces, step edges and SDs is strongly influenced by the small energy gap at the $L$ TRIM point. This small energy gap positions Bi in the vicinity of a topological phase transition into a strong TI with non-vanishing weak indices. The surface and edge states that disperse between the corresponding projections of the bulk $T$ and $L$ TRIM points, on the $\mathbb{Z}_2$ trivial side of the phase transition, are the precursors of the helical boundaries modes that would be firmly established beyond the critical transition point (for instance under application of strain or Sb doping). Indeed, doping Bi with Sb was shown to invert the bulk energy gap at $L$, while hardly affecting the surface states dispersion around the $\overline{\Gamma}$ point \cite{Benia2015}. As such, those edge modes are approximately helical and protected from back-scattering, already on the $\mathbb{Z}_2$ trivial side of the topological phase transition. Their penetration depth, the critical length associated with the phase transition, is expected to diverge as the energy gap narrows \cite{Ishida2017, Ito2016}. Their exact topological nature is, therefore, ill-defined in a finite sample \cite{Wu2013}. A similar boundary mode evolution was indeed reported across the topological phase transition from the 3D TI BiTlSe$_2$  to a trivial insulator with the substitution of Se with S \cite{Xu2011, Sato2011, Souma2012}. Similar reasoning was also applied to the edge modes of a bilayer silicene \cite{Ezawa2012}. In both cases approximate helical boundary modes were shown to emerge on approaching the topological quantum phase transition from the trivial phase. 

In summary, we have mapped the surface and edge spectra of Bi(111). We find surface and edge modes that disperse in energy and momentum between the projections of the bulk $T$ and $L$ TRIM points. The edge mode withstands crystallographic irregularities, without showing any signs of backscattering and gapping. In particular, the edge mode seems to bind to bulk SD as expected for TI with non-vanishing weak indices. These observations disfavor the recent identification of the edge mode as a hinge mode of a higher order TI class which, in the absence of protection, would be highly susceptible to hybridization by such structural perturbations. We argue that the small scale of the bulk $L$ energy gap, which in our ab initio calculation amounts to 5 meV, positions Bi within the critical region of a topological phase transition to a strong TI with non-vanishing weak indices. It also facilitates strain induced transition into the topological phase in the vicinity of the SD, as well as throughout the crystal by spurious crystallographic defects. This demonstrates that SD can serve not only as a probe for the topological classification of materials but also as a tool to manipulate and alter this classification.

\section*{Methods}
\subsection*{STM Measurements}
The STM measurements were carried out using commercial Pt-Ir tips. The tips were characterized on a freshly prepared clean Cu(111) single crystal. This ensured a stable tip with reproducible results. The dI/dV measurements were performed using the standard lock-in technique. The map parameters for Fig. \ref{fig1}f are $V=200$ mV, $I=900$ pA, $f=413$ Hz, ac rms amplitude = 1 mV. The map parameters for Fig. \ref{fig2}c and Fig. \ref{fig3}b-g are $V=200$ mV, $I=500$ pA, $f=1373$ Hz, ac rms amplitude = 3 mV.

\subsection*{Data availability}
The data that support the findings of this study are available from the corresponding author upon reasonable request.


\begin{thebibliography}{10}

\bibitem{Kane2005}
C.~L. Kane and E.~J. Mele.
\newblock Quantum spin hall effect in graphene.
\newblock {\em Phys. Rev. Lett.}, 95(22):226801, nov 2005.

\bibitem{Bernevig2006}
B.a.~Andrei Bernevig, Taylor~L Hughes, and Shou-Cheng Zhang.
\newblock Quantum spin hall effect and topological phase transition in hgte
  quantum wells.
\newblock {\em Science}, 314(5806):1757--61, dec 2006.

\bibitem{Fu2007}
Liang Fu and C~L Kane.
\newblock Topological insulators with inversion symmetry.
\newblock {\em Phys. Rev. B}, 76(4):045302, jul 2007.

\bibitem{Hasan2010}
M.~Z. Hasan and C.~L. Kane.
\newblock Colloquium: Topological insulators.
\newblock {\em Rev. Mod. Phys.}, 82(4):3045--3067, 2010.

\bibitem{Qi2011}
Xiao-Liang Qi and Shou-Cheng Zhang.
\newblock Topological insulators and superconductors.
\newblock {\em Rev. Mod. Phys.}, 83(4):1057--1110, oct 2011.

\bibitem{BERNEVIG2013}
B.~Andrei Bernevig and Taylor~L. Hughes.
\newblock {\em Topological Insulators and Topological Superconductors}.
\newblock Princeton University Press, Princeton, jan 2013.

\bibitem{Bansil2016}
A.~Bansil, Hsin Lin, and Tanmoy Das.
\newblock Colloquium : Topological band theory.
\newblock {\em Rev. Mod. Phys.}, 88(2):021004, jun 2016.

\bibitem{Wen2017}
Xiao~Gang Wen.
\newblock Colloquium: Zoo of quantum-topological phases of matter.
\newblock {\em Rev. Mod. Phys.}, 89(4), 2017.

\bibitem{Armitage2018}
N~P Armitage, E~J Mele, and Ashvin Vishwanath.
\newblock Weyl and dirac semimetals in three-dimensional solids.
\newblock {\em Rev. Mod. Phys.}, 90(1), 2018.

\bibitem{Liu1995}
Yi~Liu and Roland~E. Allen.
\newblock Electronic structure of the semimetals bi and sb.
\newblock {\em Phys. Rev. B}, 52(3):1566--1577, jul 1995.

\bibitem{Teo2008}
Jeffrey C.~Y. Teo, Liang Fu, and C.~L. Kane.
\newblock Surface states and topological invariants in three-dimensional
  topological insulators: Application to {Bi$_{1-x}$Sb$_x$}.
\newblock {\em Phys. Rev. B}, 78(4):045426, jul 2008.

\bibitem{Ohtsubo2016}
Yoshiyuki Ohtsubo and Shin~Ichi Kimura.
\newblock Topological phase transition of single-crystal bi based on empirical
  tight-binding calculations.
\newblock {\em New J. Phys.}, 18(12):123015, dec 2016.

\bibitem{Schindler2018a}
Frank Schindler, Zhijun Wang, Maia~G Vergniory, Ashley~M Cook, Anil Murani,
  Shamashis Sengupta, Alik~Yu Kasumov, Richard Deblock, Sangjun Jeon, Ilya
  Drozdov, H{\'{e}}l{\`{e}}ne Bouchiat, Sophie Gu{\'{e}}ron, Ali Yazdani,
  B~Andrei Bernevig, and Titus Neupert.
\newblock Higher-order topology in bismuth.
\newblock {\em Nat. Phys.}, jul 2018.

\bibitem{Ran2009}
Ying Ran, Yi~Zhang, and Ashvin Vishwanath.
\newblock One-dimensional topologically protected modes in topological
  insulators with lattice dislocations.
\newblock {\em Nat. Phys.}, 5(4):298--303, 2009.

\bibitem{Teo2010a}
Jeffrey C~Y Teo and C~L Kane.
\newblock Topological defects and gapless modes in insulators and
  superconductors.
\newblock {\em Phys. Rev. B}, 82(11), 2010.

\bibitem{Teo2017}
Jeffrey~C.Y. Teo and Taylor~L. Hughes.
\newblock Topological defects in symmetry-protected topological phases.
\newblock {\em Annual Review of Condensed Matter Physics}, 8(1):211--237, mar
  2017.

\bibitem{Teo2010}
Jeffrey~C.Y. Teo and C~L Kane.
\newblock Majorana fermions and non-abelian statistics in three dimensions.
\newblock {\em Phys. Rev. Lett.}, 104(4), 2010.

\bibitem{Barkeshli2012}
Maissam Barkeshli and Xiao~Liang Qi.
\newblock Topological nematic states and non-abelian lattice dislocations.
\newblock {\em Physical Review X}, 2(3), 2012.

\bibitem{Hughes2013}
Taylor~L Hughes, Hong Yao, and Xiao-Liang Qi.
\newblock Majorana zero modes in dislocations of {Sr$_2$RuO$_4$}.
\newblock {\em Phys. Rev. B}, 90:235123, 2013.

\bibitem{Read2000}
N.~Read and Dmitry Green.
\newblock Paired states of fermions in two dimensions with breaking of parity
  and time-reversal symmetries and the fractional quantum hall effect.
\newblock {\em Phys. Rev. B}, 61(15):10267--10297, apr 2000.

\bibitem{Ivanov2001}
D.~A. Ivanov.
\newblock Non-abelian statistics of half-quantum vortices in p-wave
  superconductors.
\newblock {\em Phys. Rev. Lett.}, 86(2):268--271, 2001.

\bibitem{Kelly2012}
Anthony Kelly and Kevin~M. Knowles.
\newblock {\em Crystallography and Crystal Defects}.
\newblock John Wiley {\&} Sons, Ltd, Chichester, UK, jan 2012.

\bibitem{Behnia2007}
Kamran Behnia, Luis Balicas, and Yakov Kopelevich.
\newblock Signatures of electron fractionalization in ultraquantum bismuth.
\newblock {\em Science}, 317(5845):1729--1731, sep 2007.

\bibitem{Feldman2016}
Benjamin~E. Feldman, Mallika~T. Randeria, Andr{\'{a}}s Gyenis, Fengcheng Wu,
  Huiwen Ji, R.~J. Cava, Allan~H. MacDonald, and Ali Yazdani.
\newblock Observation of a nematic quantum hall liquid on the surface of
  bismuth.
\newblock {\em Science}, 354(6310):316--321, oct 2016.

\bibitem{Randeria2018}
Mallika~T. Randeria, Benjamin~E. Feldman, Fengcheng Wu, Hao Ding, Andr{\'{a}}s
  Gyenis, Huiwen Ji, R.~J. Cava, Allan~H. MacDonald, and Ali Yazdani.
\newblock Ferroelectric quantum hall phase revealed by visualizing landau level
  wavefunction interference.
\newblock {\em Nat. Phys.}, page~1, may 2018.

\bibitem{Prakash2017}
Om~Prakash, Anil Kumar, A~Thamizhavel, and S~Ramakrishnan.
\newblock Evidence for bulk superconductivity in pure bismuth single crystals
  at ambient pressure.
\newblock {\em Science}, 355(6320):52--55, jan 2017.

\bibitem{Murakami2007}
Shuichi Murakami.
\newblock Phase transition between the quantum spin hall and insulator phases
  in 3d: Emergence of a topological gapless phase.
\newblock {\em New J. Phys.}, 9(9):356--356, sep 2007.

\bibitem{Hsieh2008}
D.~Hsieh, D.~Qian, L.~Wray, Y.~Xia, Y.~S. Hor, R.~J. Cava, and M.~Z. Hasan.
\newblock A topological dirac insulator in a quantum spin hall phase.
\newblock {\em Nature}, 452(7190):970--974, apr 2008.

\bibitem{Roushan2009}
Pedram Roushan, Jungpil Seo, Colin~V Parker, Y~S Hor, D~Hsieh, Dong Qian,
  Anthony Richardella, M~Z Hasan, R~J Cava, and Ali Yazdani.
\newblock Topological surface states protected from backscattering by chiral
  spin texture.
\newblock {\em Nature}, 460(7259):1106--1109, 2009.

\bibitem{Benia2015}
Hadj~M. Benia, Carola Stra{\ss}er, Klaus Kern, and Christian~R. Ast.
\newblock Surface band structure of {Bi$_{1-x}$Sb$_x$}(111).
\newblock {\em Phys. Rev. B}, 91(16):161406, dec 2015.

\bibitem{Zhang2009a}
Haijun Zhang, Chao~Xing Liu, Xiao~Liang Qi, Xi~Dai, Zhong Fang, and Shou~Cheng
  Zhang.
\newblock Topological insulators in {Bi$_2$Se$_3$}, {Bi$_2$Te$_3$} and
  {Sb$_2$Te$_3$} with a single dirac cone on the surface.
\newblock {\em Nat. Phys.}, 5(6):438--442, jun 2009.

\bibitem{Chen2009}
Y~L Chen, J~G Analytis, J-H Chu, Z~K Liu, S-K Mo, X~L Qi, H.~J. Zhang, D~H Lu,
  X~Dai, Z~Fang, S~C Zhang, I~R Fisher, Z~Hussain, and Z-X Shen.
\newblock Experimental realization of a three-dimensional topological
  insulator, {Bi$_2$Te$_3$}.
\newblock {\em Science}, 325(5937):178--181, 2009.

\bibitem{Hsieh2009a}
D.~Hsieh, Y.~Xia, D.~Qian, L.~Wray, J.~H. Dil, F.~Meier, J.~Osterwalder,
  L.~Patthey, J.~G. Checkelsky, N.~P. Ong, A.~V. Fedorov, H.~Lin, A.~Bansil,
  D.~Grauer, Y.~S. Hor, R.~J. Cava, and M.~Z. Hasan.
\newblock A tunable topological insulator in the spin helical dirac transport
  regime.
\newblock {\em Nature}, 460(7259):1101--1105, aug 2009.

\bibitem{Xia2009}
Y.~Xia, D.~Qian, D.~Hsieh, L.~Wray, A.~Pal, H.~Lin, A.~Bansil, D.~Grauer, Y.~S.
  Hor, R.~J. Cava, and M.~Z. Hasan.
\newblock Observation of a large-gap topological-insulator class with a single
  dirac cone on the surface.
\newblock {\em Nat. Phys.}, 5(6):398--402, jun 2009.

\bibitem{Zhang2009}
Hai~Jun Zhang, Chao~Xing Liu, Xiao~Liang Qi, Xiao~Yu Deng, Xi~Dai, Shou~Cheng
  Zhang, and Zhong Fang.
\newblock Electronic structures and surface states of the topological insulator
  {Bi$_{1-x}$Sb$_x$}.
\newblock {\em Phys. Rev. B}, 80(8), 2009.

\bibitem{Yang2012}
Fang Yang, Lin Miao, Z.~F. Wang, Meng~Yu Yao, Fengfeng Zhu, Y~R Song, Mei~Xiao
  Wang, Jin~Peng Xu, Alexei~V Fedorov, Z~Sun, G~B Zhang, Canhua Liu, Feng Liu,
  Dong Qian, C~L Gao, and Jin~Feng Jia.
\newblock Spatial and energy distribution of topological edge states in single
  bi(111) bilayer.
\newblock {\em Phys. Rev. Lett.}, 109(1), 2012.

\bibitem{Ohtsubo2013}
Yoshiyuki Ohtsubo, Luca Perfetti, Mark~Oliver Goerbig, Patrick~Le F{\`{e}}vre,
  Fran{\c{c}}ois Bertran, and Amina Taleb-Ibrahimi.
\newblock Non-trivial surface-band dispersion on bi(111).
\newblock {\em New J. Phys.}, 15(3):033041, mar 2013.

\bibitem{Drozdov2014}
Ilya~K. Drozdov, A.~Alexandradinata, Sangjun Jeon, Stevan Nadj-Perge, Huiwen
  Ji, R.~J. Cava, B.~{Andrei Bernevig}, and Ali Yazdani.
\newblock One-dimensional topological edge states of bismuth bilayers.
\newblock {\em Nat. Phys.}, 10(9):664--669, aug 2014.

\bibitem{Peng2018a}
Lang Peng, Jing-Jing Xian, Peizhe Tang, Angel Rubio, Shou-Cheng Zhang, Wenhao
  Zhang, and Ying-Shuang Fu.
\newblock Visualizing topological edge states of single and double bilayer bi
  supported on multibilayer bi(111) films.
\newblock {\em Phys. Rev. B}, 98(24):245108, 2018.

\bibitem{Ito2016}
S~Ito, B~Feng, M~Arita, A~Takayama, R.~Y. Liu, T~Someya, W.~C. Chen, T~Iimori,
  H~Namatame, M~Taniguchi, C.~M. Cheng, S.~J. Tang, F~Komori, K~Kobayashi,
  T.~C. Chiang, and I~Matsuda.
\newblock Proving nontrivial topology of pure bismuth by quantum confinement.
\newblock {\em Phys. Rev. Lett.}, 117(23), 2016.

\bibitem{Murakami2006}
Shuichi Murakami.
\newblock Quantum spin hall effect and enhanced magnetic response by spin-orbit
  coupling.
\newblock {\em Phys. Rev. Lett.}, 97(23):236805, dec 2006.

\bibitem{Kawakami2015}
Naoya Kawakami, Chun~Liang Lin, Maki Kawai, Ryuichi Arafune, and Noriaki
  Takagi.
\newblock One-dimensional edge state of bi thin film grown on si(111).
\newblock {\em Appl. Phys. Lett.}, 107(3):31602--22424, 2015.

\bibitem{Takayama2015}
A~Takayama, T~Sato, S~Souma, T~Oguchi, and T~Takahashi.
\newblock One-dimensional edge states with giant spin splitting in a bismuth
  thin film.
\newblock {\em Phys. Rev. Lett.}, 114(6), 2015.

\bibitem{Liu2017a}
Xiaogang Liu, Hongjian Du, Jufeng Wang, Mingyang Tian, Xia Sun, and Bing Wang.
\newblock Resolving the one-dimensional singularity edge states of bi(111) thin
  films.
\newblock {\em J. Phys. Condens. Matter}, 29(18):185002, may 2017.

\bibitem{Yeom2016}
Han~Woong Yeom, Kyung~Hwan Jin, and Seung~Hoon Jhi.
\newblock Topological fate of edge states of single bi bilayer on bi(111).
\newblock {\em Phys. Rev. B}, 93(7), 2016.

\bibitem{Xu2015d}
Su~Yang Xu, Madhab Neupane, Ilya Belopolski, Chang Liu, Nasser Alidoust, Guang
  Bian, Shuang Jia, Gabriel Landolt, Batosz Slomski, J.~Hugo Dil, Pavel~P.
  Shibayev, Susmita Basak, Tay~Rong Chang, Horng~Tay Jeng, Robert~J. Cava, Hsin
  Lin, Arun Bansil, and M.~Zahid Hasan.
\newblock Unconventional transformation of spin dirac phase across a
  topological quantum phase transition.
\newblock {\em Nat. Commun.}, 6(1):6870, dec 2015.

\bibitem{Barriga2018}
J~S{\'{a}}nchez-Barriga, I~Aguilera, L~V Yashina, D~Y Tsukanova, F~Freyse, A~N
  Chaika, C~Callaert, A~M Abakumov, J~Hadermann, A~Varykhalov, E~D~L Rienks,
  G~Bihlmayer, S~Bl{\"{u}}gel, and O~Rader.
\newblock Anomalous behavior of the electronic structure of
  {(Bi$_{1-x}$In$_x$)$_2$Se$_3$}.
\newblock {\em Phys. Rev. B}, 98(23):235110, dec 2018.

\bibitem{Shekhar2018}
Chandra Shekhar, Nitesh Kumar, V~Grinenko, Sanjay Singh, R~Sarkar, H~Luetkens,
  Shu-Chun Wu, Yang Zhang, Alexander~C Komarek, Erik Kampert, Yurii Skourski,
  Jochen Wosnitza, Walter Schnelle, Alix McCollam, Uli Zeitler, J{\"{u}}rgen
  K{\"{u}}bler, Binghai Yan, H-H Klauss, S~S~P Parkin, and C~Felser.
\newblock Anomalous hall effect in weyl semimetal half-heusler compounds rptbi
  (r = gd and nd).
\newblock {\em Proc. Natl. Acad. Sci.}, 115(37):201810842, sep 2018.

\bibitem{Ast2001}
Christian~R Ast and Hartmut H{\"{o}}chst.
\newblock Fermi surface of bi(111) measured by photoemission spectroscopy.
\newblock {\em Phys. Rev. Lett.}, 87(17):20, 2001.

\bibitem{Kim2014}
Sung~Hwan Kim, Kyung~Hwan Jin, Joonbum Park, Jun~Sung Kim, Seung~Hoon Jhi,
  Tae~Hwan Kim, and Han~Woong Yeom.
\newblock Edge and interfacial states in a two-dimensional topological
  insulator: Bi(111) bilayer on {Bi$_2$Te$_2$Se}.
\newblock {\em Phys. Rev. B}, 89(15), 2014.

\bibitem{Queiroz2018}
Raquel Queiroz, Ion~Cosma Fulga, Nurit Avraham, Haim Beidenkopf, and Jennifer
  Cano.
\newblock Partial lattice defects in higher order topological insulators.
\newblock {\em arXiv preprint arXiv:1809.03518}, 2018.

\bibitem{Aguilera2015}
Irene Aguilera, Christoph Friedrich, and Stefan Bl{\"{u}}gel.
\newblock Electronic phase transitions of bismuth under strain from
  relativistic self-consistent gw calculations.
\newblock {\em Phys. Rev. B}, 91(12):125129, 2015.

\bibitem{Hirahara2012}
T~Hirahara, N~Fukui, T~Shirasawa, M~Yamada, M~Aitani, H~Miyazaki, M~Matsunami,
  S~Kimura, T~Takahashi, S~Hasegawa, and K~Kobayashi.
\newblock Atomic and electronic structure of ultrathin bi(111) films grown on
  {Bi$_2$Te$_2$}(111) substrates: Evidence for a strain-induced topological
  phase transition.
\newblock {\em Phys. Rev. Lett.}, 109(22), 2012.

\bibitem{Ishida2017}
H.~Ishida.
\newblock Decay length of surface-state wave functions on bi(111).
\newblock {\em J. Phys. Condens. Matter}, 29(1), 2017.

\bibitem{Wu2013}
Liang Wu, M.~Brahlek, R.~{Vald{\'{e}}s Aguilar}, A.~V. Stier, C.~M. Morris,
  Y.~Lubashevsky, L.~S. Bilbro, N.~Bansal, S.~Oh, and N.~P. Armitage.
\newblock A sudden collapse in the transport lifetime across the topological
  phase transition in {(Bi$_{1-x}$In$_x$)$_2$Se$_3$}.
\newblock {\em Nat. Phys.}, 9(7):410--414, jul 2013.

\bibitem{Xu2011}
Su~Yang Xu, Y.~Xia, L.~A. Wray, S.~Jia, F.~Meier, J.~H. Dil, J.~Osterwalder,
  B.~Slomski, A.~Bansil, H.~Lin, R.~J. Cava, and M.~Z. Hasan.
\newblock Topological phase transition and texture inversion in a tunable
  topological insulator.
\newblock {\em Science}, 332(6029):560--564, 2011.

\bibitem{Sato2011}
T.~Sato, Kouji Segawa, K.~Kosaka, S.~Souma, K.~Nakayama, K.~Eto, T.~Minami,
  Yoichi Ando, and T.~Takahashi.
\newblock Unexpected mass acquisition of dirac fermions at the quantum phase
  transition of a topological insulator.
\newblock {\em Nat. Phys.}, 7(11):840--844, nov 2011.

\bibitem{Souma2012}
S~Souma, M~Komatsu, M~Nomura, T~Sato, A~Takayama, T~Takahashi, K~Eto, Kouji
  Segawa, and Yoichi Ando.
\newblock Spin polarization of gapped dirac surface states near the topological
  phase transition in {TlBi(S$_{1-x}$Se$_x$)$_2$}.
\newblock {\em Phys. Rev. Lett.}, 109(18):186804, 2012.

\bibitem{Ezawa2012}
Motohiko Ezawa.
\newblock Quasi-topological insulator and trigonal warping in gated bilayer
  silicene.
\newblock {\em J. Phys. Soc. Japan}, 81(10), 2012.

\bibitem{Canfield1991}
P.~C. Canfield, J.~D. Thompson, W.~P. Beyermann, A.~Lacerda, M.~F. Hundley,
  E.~Peterson, Z.~Fisk, and H.~R. Ott.
\newblock Magnetism and heavy fermion-like behavior in the rbipt series.
\newblock {\em J. Appl. Phys.}, 70(10):5800--5802, 1991.

\end{thebibliography}

\clearpage
\section*{Supplementary Information}
\setcounter{figure}{0}
\renewcommand{\thefigure}{S\arabic{figure}}
\setcounter{section}{0}
\renewcommand{\thesection}{S\arabic{section}} 

\section{Crystal growth}
\label{sec:material}
Single crystals of GdPtBi were grown by the solution growth method from Bi flux in Ta tube. The extra Bi flux was separated from the crystals by decanting the Ta tube well above the melting point of Bi. The complete procedure of crystal growth is described in the reference \cite{Shekhar2018}. In this procedure, the majority of the crystals were found at the bottom of the crucible where they stick. The grown crystals were removed with the help of a stainless steel knife after cutting the tube for further characterizations. After a careful study, we realized that Bi inclusions are invariably present in the tenth of millimeter of the bottom of the GdPtBi crystals.

\section{Effect of substrate magnetic order on Bi}
\label{sec:GdPtBi_mag}

\begin{figure}[ht]
\centering
\includegraphics[scale=1]{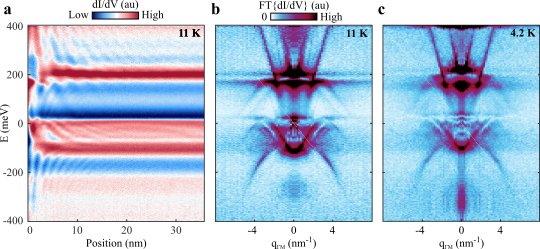}
\caption{dI/dV map measured on clean (111) surface of Bi inclusions, above and below the N\'{e}el temperature of GdPtBi, $T_N= 9.2$ K. \textbf{a,} dI/dV map measured at $T = 11\ K > T_N$ and its corresponding Fourier transform shown in \textbf{b}. \textbf{c,} Fourier transform of dI/dV map measured at $T = 4.2\ K < T_{N}$. There is no significant affect of the substrate's anti-ferromagnetic order on the surface states of bismuth.}
\label{fig:GdPtBi_mag}
\end{figure}

GdPtBi is known to be an anti-ferromagnet below its N\'{e}el transition temperature, $T_N=9.2$ K \cite{Canfield1991}. The electronic band structure of Bi remains unaffected by the magnetic ground state of the substrate, GdPtBi. This is evident from the lack of any significant difference in the QPI measured on the same region of Bi at 11 K and 4.2 K, above and below the anti-ferromagnetic transition temperature of GdPtBi, respectively (see Fig. \ref{fig:GdPtBi_mag}). The same was observed in multiple measurements across different regions of the sample.

\section{dI/dV mapping of the Screw Dislocation}

\begin{figure}[ht]
\centering
\includegraphics[scale=1.25]{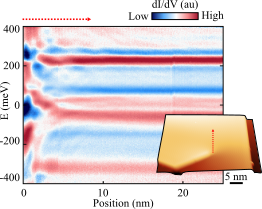}
\caption{dI/dV map measured on a Bi terrace with screw dislocation, along the red dotted line as shown in the topographic inset.}
\label{fig:LCscrew}
\end{figure}

The spectroscopic map, shown in Fig. \ref{fig:LCscrew}, was measured from the screw dislocation core, running parallel to the adjacent step edge, thus allowing us to disregard QPI effects from the adjacent step edge. Faint QPI pattern emanating from the SD is observed close to the core, consistent with the interpretation of a SD acting as a strong scatterer of impinging electrons.

\begin{figure}[ht]
\centering
\includegraphics[scale=0.75]{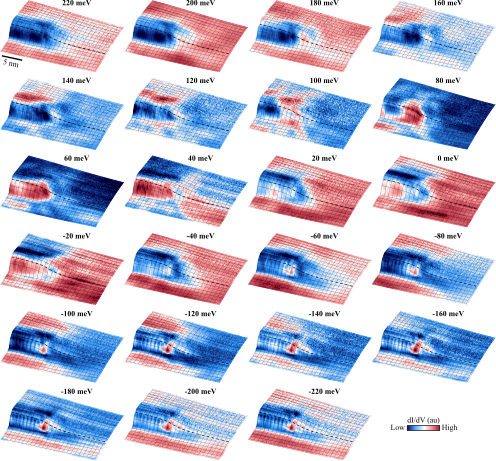}
\caption{Spectroscopic map in the vicinity of the screw dislocation, shown for various energy slices, decorated on a three-dimensionally rendered topographic image. The dashed line indicates the location of the dI/dV map shown in Fig. \ref{fig2}c.}
\label{fig:SDmap}
\end{figure}

A detailed spatially resolved spectroscopic map showing the full energy evolution of the 1D edge state associated with the screw dislocation is presented in Fig. \ref{fig:SDmap}.

\clearpage

\section{Edge states of Bi nanoribbon}

\begin{figure}[ht!]
\centering
\includegraphics[scale=1]{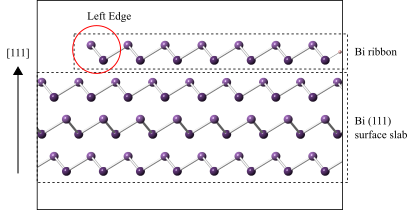}
\caption{The model to simulate Bi step edge. A ribbon structure with two zigzag edges and 14 atoms in the unit cell is placed on top of the Bi(111) surface. The 1D edge state presented in Fig. \ref{fig1}h corresponds to the density of states projected onto the left edge of the Bi nanoribbon, as indicated by the red circle. The Bi atoms on the right edge are passivated with hydrogen. The Bi surface is simulated by a slab model of 6 atomic layer thick. The large solid square is the unit cell box of the simulation.}
\label{fig:Bi_ribbon}
\end{figure}

\clearpage

\section{Model of shear strain}

\begin{figure}[ht!]
\centering
\includegraphics[scale=1]{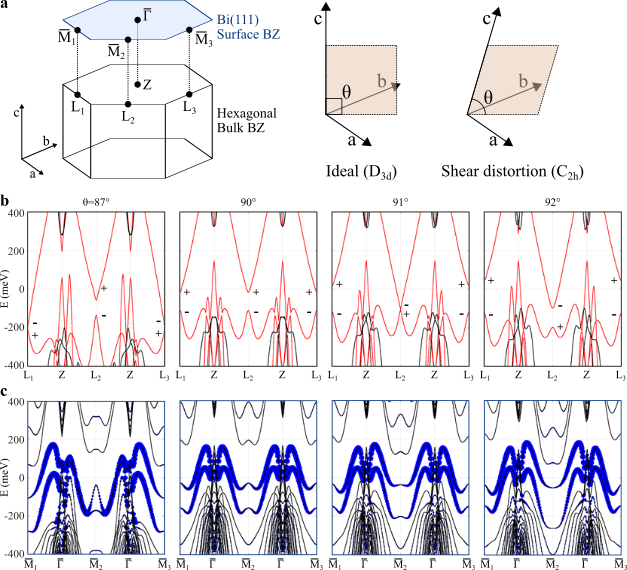}
\caption{\textbf{a,} The bulk and surface Brillouin zone of Bi's hexagonal lattice. For the ideal lattice, the space group symmetry is $D_{3d}$ while the distorted lattice with a shear strain is reduced to $C_{2h}$. In the ideal lattice, $L_{1,2,3}$ $(\bar{M}_{1,2,3})$ are equivalent to each other in the bulk (surface), while $L_{2}$ $(\bar{M}_{2})$ is distinct from $L_{1,3}$ $(\bar{M}_{1,3})$ in the distorted lattice. \textbf{b,} The bulk and \textbf{c,} surface band structures for different shear strains calculated with generalized gradient approximation. On reducing $\theta$, the energy gap shrinks at $L_{1,3}$ but increases at $L_2$. $\theta=90^\circ$ is a $\mathbb{Z}_2$ trivial phase (0;000) by counting the parities. $\theta=87^\circ$ represents a weak topological insulator (0;110) and $\theta=91^\circ$ or $92^\circ$ represents a strong topological insulator (1;111).}
\label{fig:DFTstrain}
\end{figure}

\end{document}